# Key Concepts and Principles of Blockchain Technology

Mohsen Ghorbian[1], Mostafa Ghobaei-Arani*[1]

[1] Department of Computer Engineering, Qom Branch, Islamic Azad University, Qom, Iran

*Corresponding Author email: mo.ghobaei@iau.ac.ir

**Abstract** In recent years, blockchain technology has been recognized as a transformative innovation in the tech world, and it has quickly become the core infrastructure of digital currencies such as Bitcoin and an important tool in various industries. This technology facilitates the recording and tracking of transactions across a vast network of computers by providing a distributed and decentralized ledger. Blockchain's decentralized structure significantly enhances security and transparency and prevents a single entity from dominating the network. This chapter examines blockchain's advantages, disadvantages, and applications in various industries and analyzes the implementation environments and reasons for using this technology. Also, this chapter discusses challenges such as scalability and high energy consumption that inhibit the expansion of this technology and examines blockchain technology's role in increasing efficiency and security in economic and social interactions. Finally, a comprehensive conclusion of blockchain applications and challenges has been presented by comparing blockchain applications in various industries and analyzing future trends.

# 1. Introduction

In the world of digital currencies such as Bitcoin, blockchain technology has rapidly become one of the critical infrastructures. In essence, this technology is a decentralized, distributed ledger used to record transactions across a network of computers [1]. Hence, transaction records are immutable using blockchain technology without network verification and universal changes. One of the prominent features of blockchain is its decentralized structure. Unlike traditional financial systems that depend on a central entity, blockchain operates on a peer-to-peer network where all participants or nodes have equal rights and responsibilities. This decentralization means that no person or organization can control the network completely, which increases security and transparency [2,3]. Moreover, the blockchain's immutable structure includes an encrypted hash of the previous block in each block, a timer, and transaction data, all of which are included in each block. Because of this secure link between blocks, changing information in one block is impossible without changing all subsequent blocks or requiring majority agreement. There is great difficulty in maintaining the immutability of the blockchain network [4]. Immutability is very important to maintain the integrity and trustworthiness of the data stored there. Blockchain technology offers many benefits in various industries, including advanced security, greater transparency, increased efficiency and speed, and reduced costs. The high security of the blockchain is due to the use of encryption techniques and network decentralization, which makes it resistant to fraud and unauthorized manipulation [5]. Each transaction is protected using cryptography and is linked to the previous transaction, which provides an additional layer of security. Because all parties have access to the same data, increased transparency reduces the likelihood of user disputes. Furthermore, blockchain can increase transaction speed and reduce the costs and time associated with traditional banking and record-keeping methods by eliminating intermediaries, and automating manual processes in transaction verification, and reducing operating costs, which is especially attractive in industries [6,7]. Despite the enormous potential of blockchain, the technology faces several challenges that must be overcome for wider adoption. Scalability can result in a slowdown in transaction speed and an increase in costs as the number of transactions increases; thus, the size of the blockchain increases. Solutions like layer-2 protocols and sharding are being developed to address these problems. Energy consumption is also a significant challenge, as consensus mechanisms such as proof-of-work (PoW), which are used in many blockchains, including Bitcoin, require significant energy and pose a significant environmental challenge. Using less expensive consensus mechanisms, such as proof-of-stake (PoS), is a viable alternative. Another challenge is the development of legal frameworks related to blockchain technology. Regulations must be developed to protect consumers while ensuring innovation is not hindered by the growth of technology [8]. Due to blockchain's inherent transparency, privacy concerns exist; solutions such as knowledge-free proofs and confidential transactions must be developed to protect users' information [9]. This chapter aims to examine the concepts of blockchain in depth. Additionally, exploring the complex and advanced structures and mechanisms used in the blockchain will provide a better and easier understanding of these concepts. We will investigate a comprehensive and complete overview of this technology by presenting the structures and platforms of different blockchains, their challenges, and the future approaches and advancements they will undergo.

The structure of this chapter is as follows: As a starting point, the second part thoroughly examines the necessary prerequisites, such as an Overview of Blockchain Technology, Blockchain Development History, and An Introduction to Blockchain Concepts. The third part of this chapter describes Blockchain Structure, An Overview of Blocks and Chains, Hashing Mechanism, Encryption Mechanism, Consensus Mechanism, and Types of Consensus Mechanisms. The fourth part of the chapter is Smart Contracts, Operation Mechanism of Smart Contracts, and Blockchain Platform Comparison. The fifth section comprehensively examines privacy and security in blockchain, as well as challenges and security solutions. The sixth section examines Types of Hyperledger Blockchains. The seventh section examines the Challenges and Future of Blockchain, Blockchain Adoption Challenges, and Blockchain Predictions and Trends. As the last step, the eighth section concludes this explanation.

## 2. Overview of Blockchain Technology

This section provides a comprehensive overview of blockchain technology, which explains its history, development, and applications.

### 2.1 Blockchain Development History

Blockchain technology's foundations were laid in the early 1990s, but its significant development began in 2008 with the emergence of Bitcoin. Stuart Haber and W. Scott Stornetta developed a cryptographic system in 1991 to ensure the authenticity and integrity of digital documents. Using a cryptographic hash as a timestamp chain for each digital document minimized the possibility that the information would be altered or distorted. This concept is known as the background for blockchain technology [10]. In 2008, an individual or group under the pseudonym Satoshi Nakamoto published an article promoting Bitcoin as a peer-to-peer electronic payment system using blockchain as the underlying technology. As a decentralized public ledger, the Bitcoin blockchain records and stores all Bitcoin transactions. Blocks are a series of transactions chained together and secured by cryptographic algorithms. Hence, using this technology, transactions could be conducted without a central intermediary, such as a bank, and they were also more resistant to manipulation and fraud. After Bitcoin's success, blockchain concepts also spread to other fields [11]. Ethereum, introduced in 2015 by Vitalik Buterin, is a blockchain platform that enables the execution of smart contracts. Smart contracts are self-executing computer programs that automatically execute terms and rules set by users. The technology can transform many industries, including finance, healthcare, supply chain, and voting [12]. Additionally, several other blockchain-related projects were formed, including Hyperledger, which was launched by the Linux Foundation in 2016 as a platform for developing enterprise blockchains. It consists of several projects, including Hyperledger Fabric, Hyperledger Sawtooth, and Hyperledger Indy, each with features and applications. Over time, blockchain technology gained more acceptance and has been used in various projects worldwide. Among its advantages are the increase in transparency, the increased level of security, the reduction of costs, and the enhancement of efficiency in business processes. Blockchain technology is continuing to evolve and is expected to play a more significant role in various industries and the global economy shortly [13].

## 2.2 An Introduction to Blockchain Concepts

Blockchain technology is a revolutionary technology used as the primary infrastructure for many digital currencies, including Bitcoin, which has recently gained widespread attention. In this technology, transactions and data are stored in the form of blocks, which are distributed and decentralized [14]. A blockchain is a collection of transactions that, after being confirmed by the network, are connected to a chain of previous blocks, thus the name "blockchain." The distinguishing feature of blockchain is that it is almost impossible to change the information recorded in the earlier blocks since each block contains the previous block's hash, which serves as a unique identifier for each block. This feature ensures high levels of security and transparency for blockchains [15]. The concept of decentralization means no individual or entity has complete control over a network, and all network members are responsible for verifying and recording transactions. In addition, blockchain systems are transparent and publicly visible. However, their details are preserved, enabling users to maintain privacy [16]. This feature makes blockchain systems more secure and resistant to fraud and manipulation. Therefore, blockchain technology contributes to the security and transparency of financial transactions and increases trust and transparency between financial and non-financial institutions. However, despite blockchain's high potential, several challenges must be addressed. Among these challenges are scalability and high energy consumption. The need to verify transactions by many nodes may pose scalability issues for blockchain networks [17]. Moreover, transaction verification processes, particularly in PoW networks, require significant energy consumption, which can negatively impact the environment. Hence, solutions must be developed to improve blockchain scalability and reduce energy consumption. Finally, blockchain is a revolutionary technology that may significantly impact economic and social interactions. As a transformational technology, it can become a powerful tool to increase power, security, and efficiency in various systems and organizations [18].

## 3. Blockchain Structure

This section describes the blockchain structure and consensus mechanisms. Hence, the first step is to describe the blockchain structure and its components, including the blocks, hashes, and mining process. Then, it discusses various consensus mechanisms, including PoW,PoS, etc.

### 3.1 An Overview of Blocks and Chains

The blockchain structure consists of two main components: blocks and chains. Blockchains are data units that contain transactions and related information, and each block includes a set of transactions. These blocks are linked sequentially to form a continuous chain of information. Each block is divided into three parts: transaction data, block hash for the current block, and block hash for the previous block. Transaction data contains information about all transactions conducted in the network. It is important to note that the current block hash is a unique identifier generated by a cryptographic algorithm that reveals all the information in the current block. The hash of the previous block in the chain is also an identifier and establishes the link between the last and current blocks. The hash in blocks plays a very important role in maintaining the security and integrity of a blockchain. As each block contains the previous block's hash, any changes to that block will affect the next block's hash in the chain. This feature makes it virtually impossible to manipulate information in the blockchain since changing one block requires changing all subsequent blocks, which requires a large amount of computing power. This chain structure makes blockchain a system resistant to fraud and manipulation. The process of creating a new block is known as "mining." During mining, miners use their computing power to solve complex math problems to create a new block. When a miner succeeds in solving the problem, the new block is added to the network, and all nodes in the network confirm the change. As soon as a new block has been confirmed, it is connected to the chain of blocks, and the transactions within that block are recorded as confirmed. Verifying and creating new blocks ensures that the blockchain remains distributed and decentralized. As a distributed ledger, blockchain involves all network nodes verifying and storing transactions. Each node maintains a complete copy of the blockchain and independently verifies transactions. Utilizing this feature makes the blockchain system resistant to attacks and security breaches since any changes to one node require changes to all nodes in the network. As such, blockchain is widely recognized as a highly secure and transparent system that can be used for various applications, including digital currencies, smart contracts, and supply chain management. The concepts of hashing and encryption protect data from unauthorized access in information security. Despite their similarities, these concepts serve different purposes and uses. The following will examine these two cases in detail [19].

### 3.1.1 Hashing Mechanism

The hashing mechanism is a mathematical process that converts variable-size input data into a fixed-size hash value. It is important to note that this hash value is generated uniquely for each input, so even small changes in the input can significantly influence the hash value. Hash functions are used to ensure the integrity of data, store passwords securely, and authenticate messages. One of the important characteristics of hash functions is that they are not reversible; this means that the original data cannot be retrieved from the hash value. Common examples of hash functions include MD5, SHA-1, and SHA-256. In addition to being used for the secure storage of passwords, hash functions are also important for error detection and data integrity. For example, hashing is used in network protocols to ensure that the data is not altered during transmission. Furthermore, it is used in more advanced cryptographic techniques such as digital signatures and hash-based authentication. Hash functions must be designed to resist collisions (two different inputs producing the same hash), as the presence of such collisions can lead to security weaknesses [20]. The hashing mechanism is shown in Figure 1. Hashing is commonly used to validate data and securely store passwords. The main features of hashing include the following:

- **Deterministic**: Each specific input always produces a specific hash value.
- **Efficient**: Generating the hash should be fast and efficient.
- **Pre-image resistance**: It should be infeasible to derive the original input from the hash value.
- **Collision resistance**: The probability that two different inputs produce the same hash should be very low.

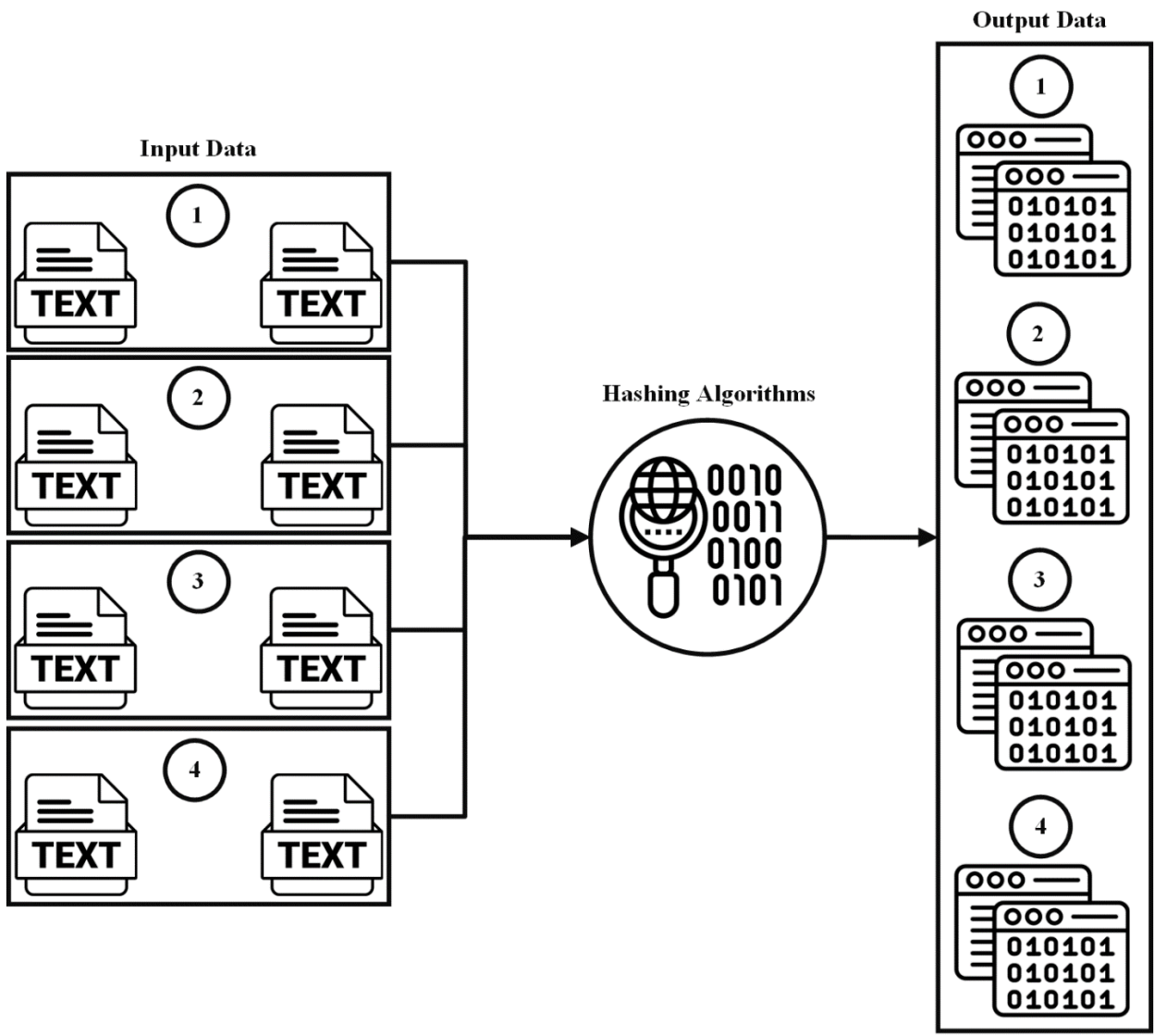

**Figure1**. Hashing Mechanism

### 3.1.2 Encryption Mechanism

Encryption mechanism refers to converting readable data (plaintext) into unreadable data (ciphertext) using cryptographic algorithms and keys to prevent unauthorized access to information. There are two main types of encryption: symmetric encryption and asymmetric encryption. In symmetric cryptography, a single key is used for encryption and decryption operations, while in asymmetric cryptography, two public and private keys are employed. The most popular encryption algorithms include AES (symmetric encryption) and RSA (asymmetric encryption). Among the many applications of cryptography are the secure transmission of information over the Internet, the protection of sensitive data such as financial and personal information, and the protection of privacy in communications. Cryptography plays a key role in establishing and maintaining trust in digital systems. For example, security protocols such as TLS/SSL, which protect web communications, rely on symmetric and asymmetric encryption techniques. In addition to providing security and authentication to digital transactions and electronic communications, cryptography is central to digital signatures and public critical infrastructures (PKI) [21]. The encryption mechanism is shown in Figure 2. Although encryption and hashing mechanisms appear similar, they differ significantly. Table 1 shows some of the most important differences between the two.

**Table 1**. The critical differences between hashing and encryption mechanisms

| Feature | Hashing | Encryption |
| --- | --- | --- |
| **Purpose** | • Ensuring data integrity and secure password storage | • Data confidentiality and security |
| **Reversibility** | • Irreversible (cannot derive original data from hash) | • Reversible (encrypted data can be decrypted with the key) |
| **Application** | • Secure password storage, data integrity, and message authentication | • Secure data transmission, protecting sensitive information |
| **Output Data Type** | • Output is always of a fixed size | • Output can have a variable size |
| **Common Examples** | • SHA-256, MD5 | • AES (symmetric encryption), RSA (asymmetric encryption) |
| Process | • Converts input data to a unique hash value | • Converts readable data to unreadable form using a key |
| Speed | • Fast and efficient | • May be slower than hashing, depending on the algorithm used |
| Use in Storage | • Yes (commonly used for secure password storage) | • No (encrypted data is not directly stored) |
| Use in Data Transmission | • No (not used for data transmission) | • Yes (used for secure data transmission) |
| Pre-image Resistance | • Yes (cannot easily derive the original input from the hash) | • Depends on the algorithm (but generally uses a private key for decryption) |

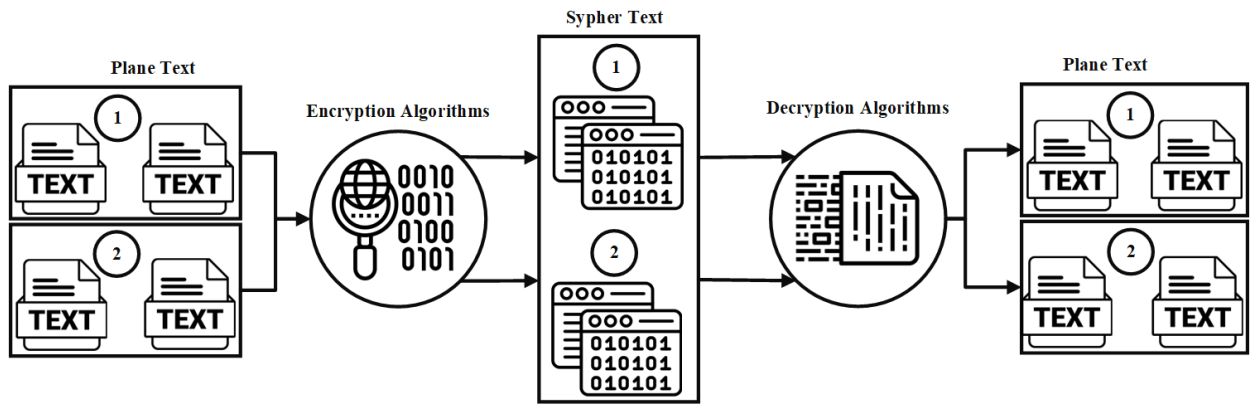

**Figure 2**. Encryption Mechanism

## 3.2 Consensus Mechanism

The consensus mechanism in distributed systems such as blockchain allows all network members to agree on the current data state. Consensus mechanisms are vital in maintaining data integrity and authenticity in decentralized networks without a central authority to manage and verify transactions. This mechanism ensures that all network participants (or nodes) agree on a standard ledger copy. In the following, some of the most critical applications of the consensus mechanism will be examined.

- **Verification and registration of transactions:** The consensus mechanism ensures that all nodes in the network confirm the authenticity of new transactions before they are added to the blockchain. Hence, this prevents fraud and double spending, as all nodes must agree on the authenticity of each transaction.

- **Network Security:** These mechanisms provide network security. Hence, the network can resist various types of attacks using consensus algorithms. When an attacker launches a 51% attack, they must control more than half of the computing power or network shares, which is very difficult in practice [22].
- **Decentralization:** Decentralization is one of the fundamental principles of blockchain technology. Consensus mechanisms allow decision-making power to be distributed throughout the network to prevent the concentration of power in a few individuals. As a result, transparency is enhanced, and power abuse is reduced.
- **Trust without intermediaries:** A trusted intermediary is needed to verify and record transactions in traditional systems. Consensus mechanisms eliminate this need and enable unknown and untrusted nodes to trust each other and validate transactions. Consequently, transaction costs are reduced, and efficiency is increased [23].

The way of functioning and how to implement the process in which the consensus mechanism leads to the confirmation of transactions and, consequently, the block is designed in such a way that all the steps are connected in a chain. Consequently, understanding how the consensus mechanism works can reduce the complications of implementing this process. Accordingly, as illustrated in Figure 3, this process usually involves several steps, which are discussed in greater detail below.

- **Block Proposal**: Nodes or miners propose a new block containing several transactions.
- **Verification and Validation**: Other nodes review and verify this block. This verification can include checking the validity of transactions and ensuring no fraud.
- **Consensus**: Nodes agree to add the new block to the blockchain. This agreement is usually achieved using the network's consensus algorithm.
- **Addition to the Blockchain**: After reaching a consensus, the new block is added to the blockchain, and the transactions are finalized

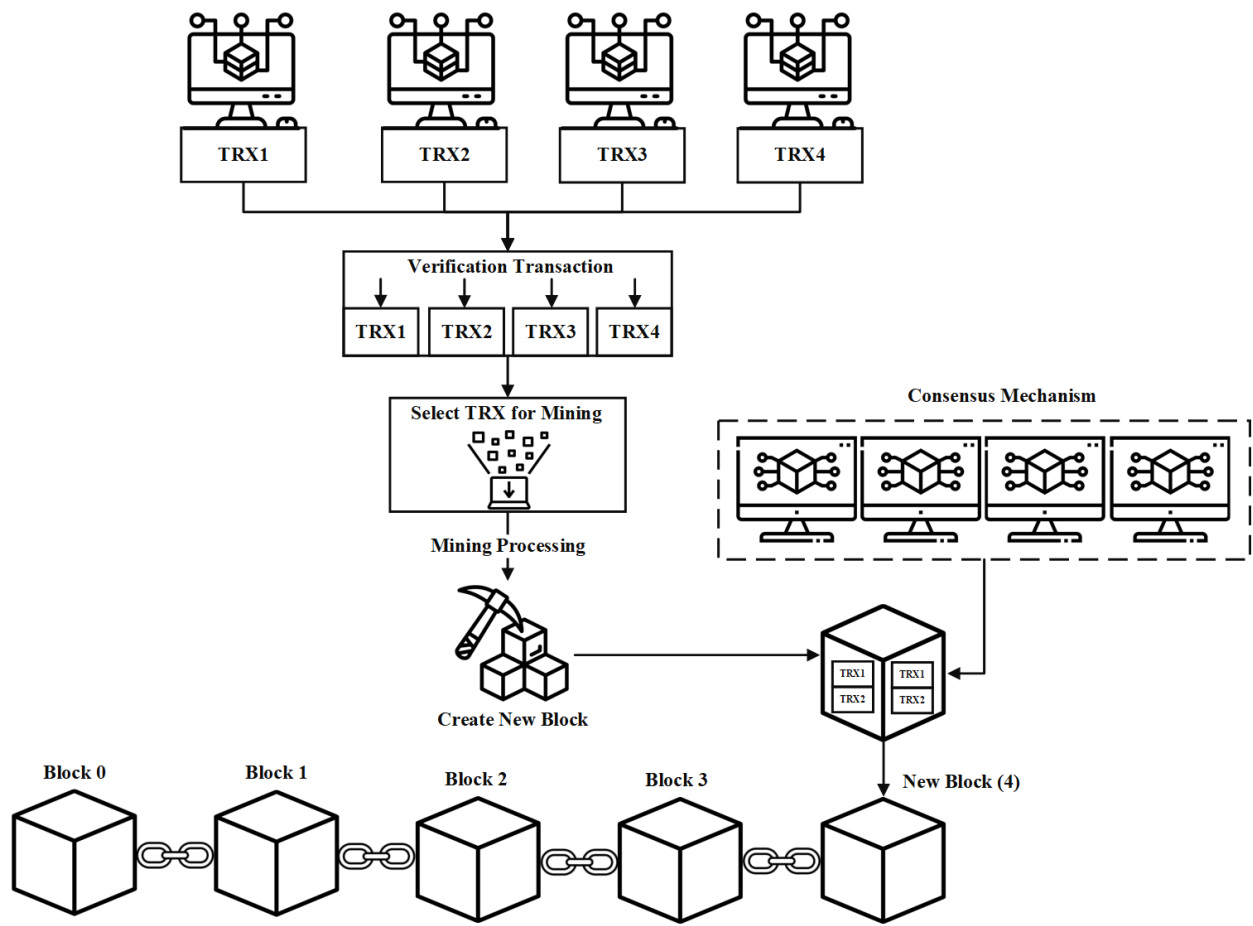

**Figure 3**. Consensus Mechanism

### 3.3 Types of Consensus Mechanisms

The blockchain's consensus mechanisms are the backbone of its security and performance. They ensure that all participants in the network are in agreement concerning the current state of the ledger and that transactions are verified in a fair and decentralized manner. Since Bitcoin introduced a PoW mechanism, various consensus models have emerged, each offering advantages such as improved efficiency, reduced energy consumption, and increased security. In this section, we tried discussing ten prominent consensus mechanisms that were reviewed, including PoW, PoS, Delegated Proof of Stake (DPoS), Proof of Capacity (PoC), Proof of Burn (PoB), Proof of History (PoH), Proof of Authority (PoA), Proof of Activity (PoA), and Proof of Elapsed Time (PoET). Each of these mechanisms plays a unique role in the evolution and improvement of blockchain networks. Types of consensus mechanisms are shown in Figure 4.

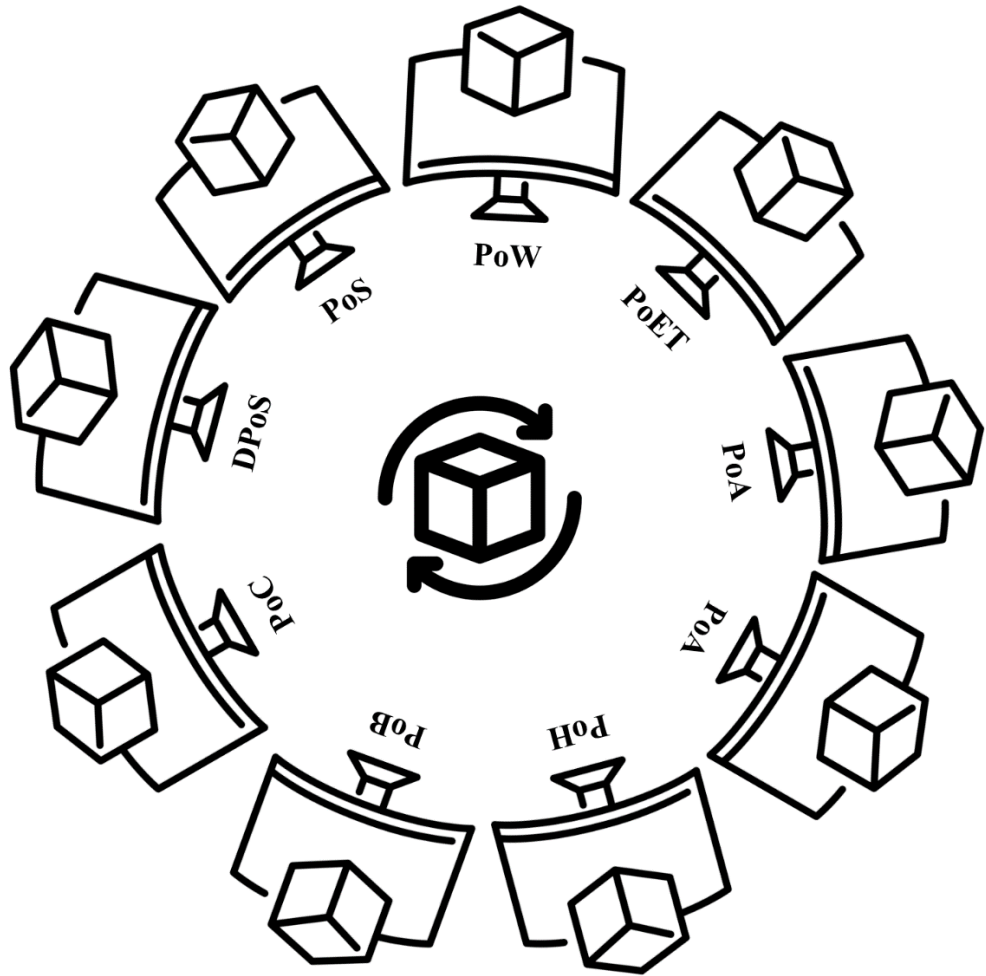

**Figure 4**. Types of Consensus Mechanisms

- **PoW:** Bitcoin introduced a PoW mechanism where miners competed against each other to solve complex mathematical problems. The first miner to solve a problem receives a reward and adds a new block to the blockchain. Although this system is highly secure, it consumes substantial energy and requires considerable time to verify transactions [24].

- **PoS:** In the PoS mechanism, blocks are validated based on the user's digital currency. In contrast to mining, users pledge their digital currency and are selected to create new blocks based on their stakes. This mechanism consumes less energy and takes less time to verify blocks [25].

- **DPoS:** In this mechanism, network users choose representatives who perform the validation process and generate blocks. These representatives are responsible for maintaining the network and are elected and controlled by user votes. DPoS is more efficient and increases user participation, but it also carries the risk of concentrating power in a limited number of hands [26].

- **PoC:** Miners mine blocks using their hard disk storage in this mechanism. Since the system relies on storing large amounts of data, it uses less energy than PoW, but it requires a large amount of storage space and takes longer to search and retrieve the data.
- **PoB:** In this mechanism, users earn the right to mine or validate blocks by "burning" or destroying their digital currencies (by sending them to an inaccessible address). The process is designed to demonstrate users' long-term commitment to the network. Although PoB consumes less energy than PoW, some users may be discouraged from participating due to the constant loss of cryptocurrency [27].
- **PoH:** A PoH mechanism records the chronological sequence of events and transactions without public consensus. This mechanism complements other consensus mechanisms, such as PoS, improves transaction confirmation times, and makes the network more scalable. PoH is used in projects such as Solana.
- **PoA:** In this mechanism, a set of well-known and valid nodes is responsible for validating and producing blocks. These nodes should usually have a specific and valid identity. PoA is suitable for private networks and consortiums and reduces transaction confirmation time, but the concentration of power in the hands of authoritative nodes may challenge decentralization [28].
- **PoA:** The PoA process comprises a PoW and a PoS system. After mining, stakeholders must verify the blocks to prove they have been mined. This mechanism aims to enhance security and decentralization while combining the advantages of both methods. However, it is more complex than simpler mechanisms.
- **PoET:** Intel primarily developed this mechanism for permission-based blockchain networks. In PoET, nodes need to wait for a random amount of time, and the first node to pass the time is entitled to create a block. Hence, using reliable hardware, this method generates random times without consuming as much energy as PoW [29].

In Table 2, different consensus mechanisms in blockchains are compared. This comparison evaluates each mechanism using five parameters: energy consumption, verification time, security, decentralization, and implementation complexity. Based on these parameters, it is possible to more accurately determine the strengths and weaknesses of each mechanism and make a more appropriate choice for the different needs of blockchain networks.

Table 2. Comparison of Blockchain Consensus Mechanisms

| Mechanism | Energy Consumption | Confirmation Time | Security | Decentralization | Implementation Complexity |
|---|---|---|---|---|---|
| PoW | • Very High | • Medium to High | • Very High | • High | • Medium |
| PoS | • Low | • Fast | • High | • Medium | • Medium |
| DPoS | • Low | • Very Fast | • High | • Medium to Low | • Medium |
| PoC | • Low | • Medium | • Medium | • High | • Medium |
| PoB | • Medium | • Medium | • High | • High | • Medium |
| PoH | • Very Low | • Very Fast | • High | • High | • High |
| PoA | • Very Low | • Very Fast | • High | • Low | • Low |
| PoA (Activity) | • Medium | • Medium | • High | • High | • High |
| PoET | • Very Low | • Fast | • High | • Medium | • High |

## 4. Smart Contracts

The term smart contracts refers to self-executing programs on the blockchain that permit transactions and agreements to be executed automatically without the involvement of intermediaries. These contracts are stored on the blockchain, which is immutable and transparent. The terms and conditions of smart contracts are written in code, and as soon as the predetermined conditions are met, the desired operation is performed. These features increase efficiency, reduce costs, and reduce the risk of human error and fraudulent transactions. One of the pioneers of this technology, Ethereum, has developed a platform for developers to run their smart contracts. Smart contracts in blockchain are immutable, extremely secure, and a powerful tool for enhancing transparency and trust in transactions [30].

### 4.1 Operation Mechanism of Smart Contracts

The operation mechanism of smart contracts is that the programming code related to the contract is stored in the blockchain and is executed when necessary. For example, suppose two parties have a contract in which one party must pay an amount and the other must deliver a product. In the smart contract code, the contract terms are stated so that the amount is automatically deposited into the other party's account upon meeting particular conditions (such as the delivery of goods or reaching a specific date). This process eliminates the need for traditional intermediaries, such as lawyers and financial institutions, and it also eliminates human error and fraud. Smart contracts are written on blockchains like Ethereum using programming languages like Solidity. In addition, smart contracts are highly secure, as they cannot be altered after they have been stored in the blockchain, thereby increasing their security. As a result of these features, smart contracts are an efficient and reliable tool for various industries, including finance, insurance, real estate, and supply chain. The applications of smart contracts in different industries are extensive and diverse. In real estate, smart contracts can automate the buying and selling processes so property ownership is transferred only after full payment. Also, smart contracts can be used in the insurance industry to check and pay claims automatically; for example, if an accident is mentioned in the insurance contract, the claim amount is automatically deposited into the policyholder's account. In the supply chain, smart contracts can assist in tracking and managing inventory, which will enable every step from production to delivery to the end customer to be documented and verified. Through these applications, many sectors will become more efficient, reduce costs, and be more transparent, leading to broader technology adoption [31]. The smart contract mechanism is shown in Figure 5.

.

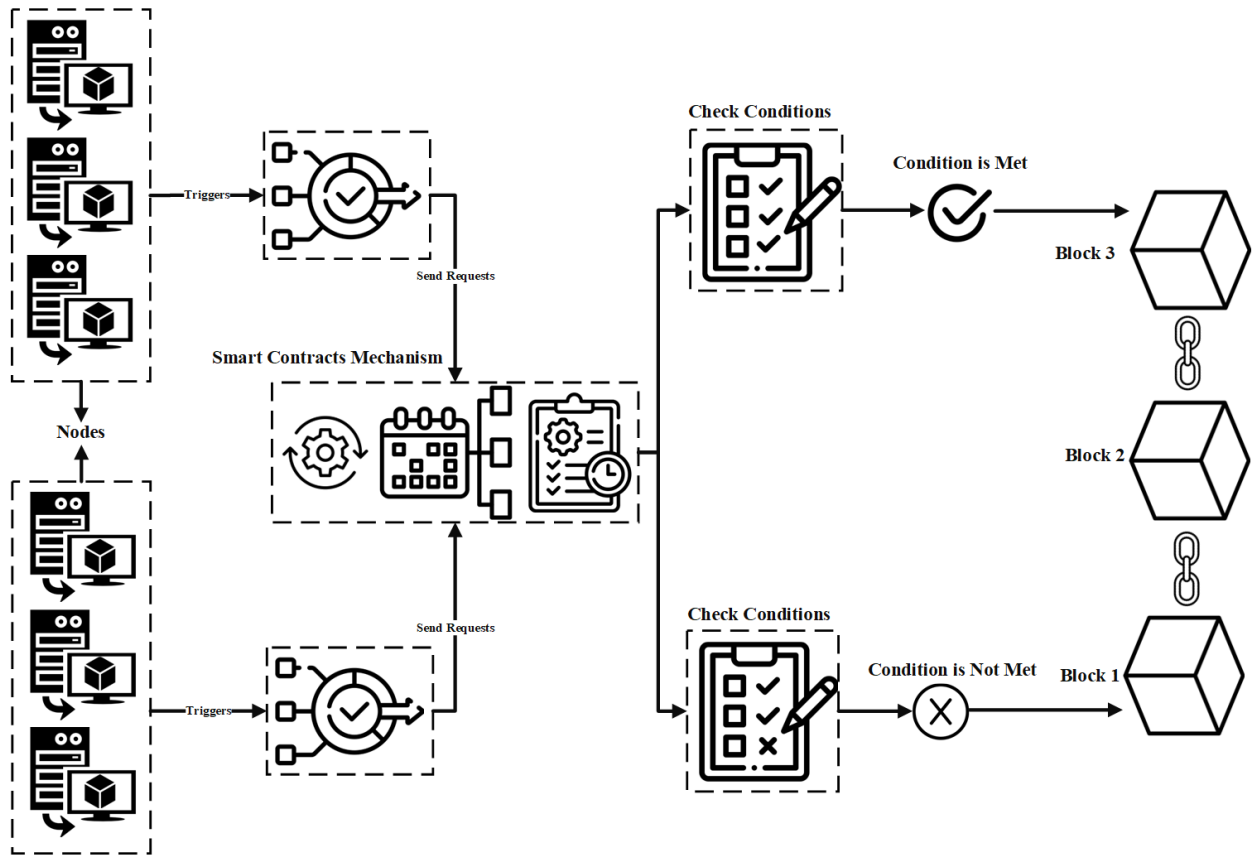

**Figure 5**. Smart Contract Mechanism

## 4.2 Blockchain Platform Comparison

Considering the expansion of the use of blockchain in various industries and the emergence of multiple platforms, it is essential to compare these technologies to select the most suitable platform to implement smart contracts. The comparison is based on six main parameters: popularity, scalability, transaction cost, security, transaction time, and programming language, as shown in Table 3. This comparison also includes permission-based and permission-less blockchains to understand their differences better. Permission-less blockchains such as Ethereum, Solana, and Cardano are ideal for open and decentralized projects due to their popularity and high security, while permission-based blockchains such as Hyperledger Fabric and Corda focus on security and efficiency for enterprise and permission-based applications that need access control and high speed, they are very suitable. Each of these parameters plays a vital role in determining the performance and capabilities of the platform and can help developers and businesses choose the best option based on their specific requirements [32].

**Table 3**. Comparison of Permission-less and Permission-based Blockchain Platforms

|  | Type of Blockchain | Popularity | Scalability | Transaction Cost | Security | Transaction Time | Programming Language |
|---|---|---|---|---|---|---|---|
| **Permission-less Blockchains** | • Ethereum | • Very High | • Medium | • Medium | • Very High | • Medium | • Solidity |
|  | • Binance Smart Chain (BSC) | • High | • High | • Low | • High | • Fast | • Solidity |
|  | • Polkadot | • High | • Very High | • Medium | • High | • Fast | • Rust<br>• Substrate |
|  | • Cardano | • High | • High | • Low | • Very High | • Fast | • Plutus<br>• Haskell |
|  | • Solana | • High | • Very High | • Very Low | • High | • Very Fast | • Rust |
| **Permissioned Blockchains** | • HLF | • High | • Very High | • Low | • Very High | • Fast | • Go<br>• Java<br>• JavaScript |
|  | • Corda | • High | • High | • Low | • Very High | • Fast | • Kotlin<br>• Java |
|  | • Quorum | • Medium | • High | • Low | • High | • Fast | • Solidity |
|  | • R3 | • Medium | • High | • Low | • Very High | • Fast | • Kotlin<br>• Java |
|  | • Avalanche | • Medium | • Very High | • Low | • Very High | • Fast | • Solidity |

According to the comparison of blockchain technologies based on smart contracts, each platform has advantages and characteristics that make it suitable for specific applications. Due to their popularity and high level of security, permission-less blockchains like Ethereum, Solana, and Cardano are ideal for open and decentralized projects. However, some of these blockchains may have difficulties with scalability and transaction costs. On the other hand, permission-based blockchains, including Hyperledger Fabric and Corda, which emphasize security and efficiency, are ideally suited for enterprise and permission-based applications requiring high speeds and access control. Each of these technologies can deliver significant productivity and efficiency in different fields depending on the specific needs of the project and the desired goals. Finally, choosing the most appropriate platform depends on the specific needs of the project and the desired goals.

## 5. Privacy and Security in Blockchain

The privacy issue in blockchain technology is a vital topic that requires special attention because of its inherent characteristics. As a distributed ledger, blockchain stores information publicly and is accessible to all network members. This transparency helps increase trust and security, but at the same time, it can create privacy issues. One of the significant challenges when using blockchains is that transactions and data are permanently and immutable and can easily be tracked and analyzed by third parties. These challenges are addressed through several techniques designed to protect users' privacy and personal information. Advanced encryption is one of these techniques. Cryptographic methods such as public and private key cryptographic algorithms increase security and privacy in blockchain. These methods ensure that only the private key owner can access and perform relevant transactions [33]. Additionally, homomorphic encryption techniques facilitate computations on encrypted data without requiring them to be decrypted. Additionally, there are methods for maintaining privacy on the blockchain besides encryption, such as Confidential Transactions, which hide the exact number of transactions so that only authorized individuals can view them. Furthermore, Stealth Addresses allow users to create one-time addresses for each transaction, making tracking and identifying transactions extremely difficult. In addition, using hybrid networks (Hybrid Networks) that combine public and private blockchains is also an effective means of maintaining privacy. This type of network stores sensitive information in private blockchains, while public blockchains contain only essential and non-sensitive data. Hence, using this approach, it is possible to balance transparency and privacy so that sensitive information about users is not accessible to the public while maintaining the transparency necessary to generate trust [34].

### 5.1 Challenges and Security Solutions

Blockchain technology has created tremendous advancements in various fields but also faces significant security challenges (e.g., penetration and hacker attacks). The challenges vary depending on the type of blockchain (public, private, consortium, and hybrid), and specific solutions are required to counter those challenges. In this regard, careful examination of security challenges and providing appropriate solutions for each type of blockchain is of particular importance to benefit from the advantages of this technology in the best way and reduce its weaknesses [35]. In Table 4, these challenges are examined in detail, as well as the proposed solutions and the type of blockchain that can address each challenge.

Table 4. Blockchain Security Issues and Solutions

| Security Issue | Solutions | Suitable Blockchain Type |
|---|---|---|
| **Double Spending Attacks** | • Advanced consensus protocols like PoS | • Public Blockchain |
| **51% Attacks** | • Widespread distribution of processing power, PoS protocols | • Public Blockchain |
| **User Privacy Breach** | • Confidential Transactions, Stealth Addresses | • Private Blockchain |
| **Cyber Attacks and Network Intrusion** | • Public and private key cryptography algorithms | • Public and Private Blockchain |
| **Security Flaws in Digital Wallets** | • User education on secure usage methods | • All types of Blockchain |
| **Transaction Tracking and Analysis** | • Advanced cryptographic techniques like homomorphic encryption, Hybrid Networks | • Hybrid Blockchain |
| **Sybil Attacks** | • Sybil-resistant consensus protocols like PoS | • Public and Consortium Blockchain |
| **DDoS Attacks** | • Rate limiting mechanisms, resource distribution | • Private and Consortium Blockchain |
| **Security Flaws in Smart Contracts** | • Thorough security audits, security analysis tools | • Public and Consortium Blockchain |
| **Private Key Misuse** | Secure key management techniques like Mult signature, Hardware Security Modules (HSM) | Public and Private Blockchain |

In comparing attacks and security solutions in the field of blockchain, blockchain type plays a significant role in addressing these issues. Advanced consensus protocols are required to address challenges such as double-spending and 51% more attacks on public blockchains. Along with privacy disclosures and cyber-attacks, public and private blockchains face cyber-attacks, which require advanced cryptographic techniques. In addition, public blockchains are more vulnerable to intelligent contracts and Sybil attacks, whereas consortium and private blockchains are also vulnerable to DDoS attacks and transaction tracking. The selection and implementation of appropriate solutions can increase the security and privacy of blockchain networks in general, depending on the type of blockchain and its associated security challenges.

## 6. Types of Hyperledger Blockchains

The Hyperledger series of blockchains, one of the most advanced blockchain technologies, has attracted much attention due to its unique characteristics and capabilities. The Hyperledger company developed these blockchains under the supervision of the Linux Foundation, and they are intended to support law enforcement programs and applications, manage operations, and record business transactions. These blockchains have a flexible architecture that allows developers to implement and manage blockchain networks based on specific organizational needs. The Hyperledger blockchain series includes several frameworks and platforms optimized for a specific application [36]. The following will examine the most important frameworks and tools for the Hyperledger series of blockchains.

- **Hyperledger Fabric (HLF)**: The Hyperledger Fabric project is a modular, flexible blockchain framework for consortium environments. This framework supports smart contracts written in a wide range of languages, including Go, Java, JavaScript, and Python, and allows users to create private channels to ensure data privacy. In addition to supply chain and banking applications, Hyperledger Fabric is also suited to smart contracts and international trade [37].
- **Hyperledger Sawtooth (HLS)**: The Hyperledger Sawtooth framework is a modular blockchain framework that emphasizes parallel transaction execution and uses the PoET consensus algorithm. These features make Sawtooth suitable for IoT, tax, and healthcare applications. This framework has a moderate to high implementation complexity and supports Python, JavaScript, Rast, and C++ programming languages [38].
- **Hyperledger Iroha (HLIr)**: The Hyperledger Iroha is an easy-to-use and quick-to-develop blockchain application framework. Due to its simple design and the use of C++ and Java programming languages, this framework is ideal for asset management and financial systems. Iroha has fewer advanced features than other Hyperledger frameworks and is better suited to simpler applications [39].

- **Hyperledger Indy (HLIn)**: The Hyperledger Indy blockchain framework manages digital identities and maintains privacy. Moreover, the framework supports Python and Rast programming languages for creating and managing independent digital identities. Indy is suitable for applications involving digital identity and authentication, and its implementation complexity is moderate [40].
- **Hyperledger Burrow (HLBu)**: The Hyperledger Burrow blockchain framework supports Ethereum Virtual Machines (EVM) and is intended to execute Solidity-based smart contracts. Due to its speed of transaction execution and parallel processing, Burrow is well-suited to financial systems and digital tokens. This framework has fewer features than Fabric, and advanced consensus algorithms are unsupported.
- **Hyperledger Besu (HLBe)**: The Hyperledger Besu is an open-source Ethereum client designed for public and private networks. This framework supports various consensus algorithms such as PoW and PoA and is suitable for enterprise applications that require compatibility with the Ethereum network. Besu uses Java programming and has a relatively simple implementation [41].

The introduced blockchains have different features and capabilities, and each is suited to use in a different industry. As illustrated in Table 5, each blockchain has different advantages, disadvantages, applications, implementation complexity, flexibility, and programming languages.

Table 5: Detailed Comparison of Hyperledger Blockchain Platforms

| Platform | Advantages | Disadvantages | Applications | Implementation Complexity | Flexibility | Programming Languages |
|---|---|---|---|---|---|---|
| HLF | • High flexibility, data privacy<br>• Smart contracts support | • High complexity<br>• Requires extensive management | • Supply chain banking<br>• Smart contracts<br>• International trade | • High | • High | • Go<br>• Java<br>• JavaScript<br>• Python |
| **HLS** | • Parallel transaction execution | • Medium to high complexity | • Internet of Things (IoT)<br>• Taxation<br>• Healthcare | • Medium to high | • Medium | • Python<br>• JavaScript<br>• C++ |
| **HLIr** | • Simple<br>• User-friendly | • Fewer advanced features | • Asset management<br>• Financial systems | • Low | • Medium | • C++<br>• Java |
| **HLIn** | • Digital identity management | • Medium complexity | • Digital identity management<br>• Authentication | • Medium | • Medium | • Python<br>• Rust |
| **HLBu** | • Solidity smart contracts<br>• High execution speed | • Fewer advanced features<br>• High complexity | • Financial systems, digital tokens | • High | • Medium | • Solidity<br>• JavaScript |
| **HLBe** | • Ethereum support | • Medium complexity | • Ethereum public and private networks | • Medium | • Medium | • Java |

Upon comparing hyper ledger projects, it is evident that each framework has its unique features and applications. HLF has high flexibility and data privacy and is suitable for supply chain and banking. Due to its parallel execution of transactions and the PoET algorithm, HLS is suitable for IoT and taxation applications. The simplicity of HLIr makes it suitable for asset management and simple financial systems. In addition to digital identity management and privacy, HLIn is also used to authenticate digital data. In addition to being able to implement smart contracts in the Solidity language, HLBu is suitable for financial systems and digital tokens. The appropriate choice between these frameworks requires carefully assessing each project's needs and conditions.

7. **Challenges and Future of Blockchain**

Blockchain is an emerging technology with unique characteristics and capabilities rapidly growing and expanding. However, scalability, privacy protection, high transaction costs, and compliance with various laws and regulations still affect its development and prosperity. The future of blockchain depends on how it can respond to these challenges and be applied in several fields, including banking, supply chain, health, governance, elections, and more. Similarly, positive developments in modern technology, such as the development of 5G networks and artificial intelligence, can lead to new areas for blockchains to be flexible and expanded.

**7.1 Blockchain Adoption Challenges**

Many technical and non-technical challenges accompany blockchain technology's widespread adoption and use. These technical challenges include:

- **Scalability**: One of the biggest challenges of blockchain is scalability. Certain blockchain networks, such as Bitcoin, have limitations to transaction processing, which prevents massive growth at larger scales. New technologies, such as more effective and optimal consensus algorithms, must be developed to address this challenge.
- **Security**: Security is a cornerstone of blockchain, but it is not immune to malicious attacks. Given the abundance of digital currencies and other sensitive data in large blockchain networks, the need to bolster security and thwart attacks is a critical vulnerability that must be addressed [42].
- **Privacy Protection**: Several blockchain systems, such as Bitcoin, do not guarantee that users' privacy will be protected. Hence, this can lead to unwanted visits and tracking for some users and businesses.
- **Transaction Costs**: Blockchain transactions can be expensive on some networks, especially if the network is subject to large, complex transactions. In some cases, these costs may present a barrier to the widespread adoption of blockchain technology [43].

To explain important cases of non-technical challenges, the following are provided:

- **Legal Settings and Regulations**: Using blockchain in numerous areas necessitates the creation of appropriate laws and regulations. These legal frameworks are crucial in making blockchain technology acceptable for business and public use, particularly in setting government tax policy.

- **Cultural Adoption and Mediation**: Implementing blockchain may require significant changes in organizational culture and work processes for some organizations. Hence, this can be one of the most important non-technical challenges, requiring employees and managers to be trained and aware of the implications [44].
- **Provision of Human Resources**: Blockchain can only be implemented and managed with the help of experienced and specialized human resources, which may present a significant challenge in certain instances.
- **Technology Change Management**: Blockchain technology is rapidly developing, and managing continuous changes and updates requires specific attention and strategies to avoid unintended consequences [45].

Several challenges stand in the way of blockchain adoption and expansion, and continuous effort is required to overcome and advance this technology into blockchain-based communities.

**7.2 Blockchain Predictions and Trends**

The future of blockchain technology has been extensively studied and analyzed and is expected to bring about significant changes in industries and the global economy. Blockchain is predicted to serve as the base infrastructure for data and value transfer in the IoT and 5G and 6G wireless networks. These two technologies allow organizations and companies to leverage the IoT more securely, faster, and with improved cross-device capabilities. Due to the need for more transparency and security in the global economy, blockchain can be more widely accepted in supply chains, banking, insurance, and e-government. In addition to improving reliability and reducing costs, smart contracts used in the blockchain can be executed automatically and operate without human intervention. In addition, blockchain can be important in changing how information and data are managed, improving efficiency, and facilitating data sharing between organizations and bodies.

**8. Conclusion**

Blockchain has been one of the most significant technological innovations in several fields, including digital currencies, supply chains, banking, health, and e-government. Hence, this technology has overcome many challenges in traditional systems by providing a distributed and decentralized ledger system. This chapter closely examines the advantages and disadvantages of blockchain and analyzes the technology's challenges, including scalability, privacy, and transaction costs. Additionally, this technology has been evaluated to transfer data and value in the IoT and on wireless networks operating at 5G and 6G speeds. By comparing blockchain applications across various industries and analyzing recent technological developments, this chapter examines the potential and limitations of blockchain technologies. Also, a specific focus is given to technological developments in combining blockchain with other technologies and emerging trends that demonstrate the effectiveness of this technology.